\PassOptionsToPackage{unicode}{hyperref}
\PassOptionsToPackage{hyphens}{url}
\PassOptionsToPackage{dvipsnames,svgnames,x11names}{xcolor}
\documentclass[
  oneside,twocolumn,eqsecnum,nofootinbib,showkeys,aps,prd]{revtex4-2}
\usepackage{amsmath,amssymb}
\usepackage{iftex}
\ifPDFTeX
  \usepackage[T1]{fontenc}
  \usepackage[latin1]{inputenc}
  \usepackage{textcomp} % provide euro and other symbols
\else % if luatex or xetex
  \usepackage{unicode-math} % this also loads fontspec
  \defaultfontfeatures{Scale=MatchLowercase}
  \defaultfontfeatures[\rmfamily]{Ligatures=TeX,Scale=1}
\fi
\usepackage{lmodern}
\ifPDFTeX\else
\fi
\IfFileExists{upquote.sty}{\usepackage{upquote}}{}
\IfFileExists{microtype.sty}{% use microtype if available
  \usepackage[]{microtype}
  \UseMicrotypeSet[protrusion]{basicmath} % disable protrusion for tt fonts
}{}
\makeatletter
\@ifundefined{KOMAClassName}{% if non-KOMA class
  \IfFileExists{parskip.sty}{%
    \usepackage{parskip}
  }{% else
    \setlength{\parindent}{0pt}
    \setlength{\parskip}{6pt plus 2pt minus 1pt}}
}{% if KOMA class
  \KOMAoptions{parskip=half}}
\makeatother
\usepackage{xcolor}
\usepackage{graphicx}
\makeatletter
\def\maxwidth{\ifdim\Gin@nat@width>\linewidth\linewidth\else\Gin@nat@width\fi}
\def\maxheight{\ifdim\Gin@nat@height>\textheight\textheight\else\Gin@nat@height\fi}
\makeatother
\makeatletter
\def\fps@figure{htbp}
\makeatother
\ifLuaTeX
\usepackage[bidi=basic]{babel}
\else
\usepackage[bidi=default]{babel}
\fi
\babelprovide[main,import]{english}
\def\languageshorthands#1{}
\numberwithin{equation}{section}
\newcommand{\mycite}[1]{\citep{#1}}
\makeatletter
\renewcommand{\eqnum}{\refstepcounter{equation}\quad\textup{\tagform@{\theequation}}}
\makeatother

\ifLuaTeX
  \usepackage{selnolig}  % disable illegal ligatures
\fi
\IfFileExists{bookmark.sty}{\usepackage{bookmark}}{\usepackage{hyperref}}
\IfFileExists{xurl.sty}{\usepackage{xurl}}{} % add URL line breaks if available
\hypersetup{
  pdflang={en},
  colorlinks=true,
  linkcolor={blue},
  filecolor={Maroon},
  citecolor={Blue},
  urlcolor={Blue},
  pdfcreator={LaTeX via pandoc}}

\date{\today}

\begin{document}

\title{Perihelion precession in non-Newtonian central potentials}
\author{Michele Andreoli}
\email{micheleandreoli@cnr.it}

\affiliation{CNR, National Research Council~\\
 Via Giuseppe Moruzzi, I-56124 Pisa, Italy}
\thanks{Accepted for publication in \emph{Astrophysics and Space Science,
SpringerNature}}
\begin{abstract}
\emph{High order corrections to the perihelion precession are obtained
in non-Newtonian central potentials, via complex analysis techniques.
The result is an exact series expansion whose terms, for a perturbation
of the form $\delta V=\frac{\gamma}{r^{s}}$, are calculated in closed
form. To validate the method, the series is applied to the specific
case of s=3, and the results are compared with those presented in
literature, which are relate to the Schwarzschild metric. As a further
test, a numerical simulation was carried out for the case where s=4.
The algebraic calculations and numerical simulations were carried
out via software with symbolic capabilities.}
\end{abstract}

\maketitle

\section{Introduction}

Precession of the perihelion is a phenomenon concerning the movement
of a planet's perihelion, which is the point in orbit where the planet
is closest to the Sun. In an elliptical orbit, the perihelion does
not remain fixed, but slowly shifts over time. This movement is caused
mainly by the gravitational influence of other bodies in the solar
system, especially the more massive planets (notably Jupiter), and
by the curvature of spacetime as predicted by Einstein's general theory
of relativity. 

The angular displacement $\Delta\phi$ can be calculated through an
integral over the planet's orbit, whose analytical form is easily
expressed in terms of an effective potential. General Relativity modifies
this effective potential by adding a \emph{correction} of the power-law
form $\delta V=\gamma/r^{s}$ with exponent $s=3$. For the case where
$s=3$, the result can be expressed both with elliptical integrals
(see, for example, \mycite{sjwalters-1}) and with power series (see,
for example, \mycite{poveda-1}). 

The interest in this type of calculation, starting from Einstein's
famous prediction regarding the anomalous behavior of Mercury's perihelion,
has never waned over the years. Corrections to the higher order could
be still useful both in the astronomical field (for example: gravitational
perturbations due to other planets; irregular shapes in the mass distribution;
new gravitational models; special solutions to the Einstein equations,
such the Zipoy-Voorhees metric, etc) that, more generally, as a theoretical
tool.

The problem with this integral is its \emph{divergence} at the points
of inversion of the motion. Landau (see \mycite{landau1} for $s=2,3$)
bypasses the problem, through an integration by parts, but this method
is applicable only to the first order in the perturbation. For higher
exponents $s$, the calculation of the inversion points and, even
more, the connection between physical quantities (such as energy $E$
and angular momentum $L$) and orbital parameters (such excentricity
and semi-axis), requires computing the roots of high-degree polynomials,
and this is unsolvable in closed form. Many authors limit themselves
to using the same formulas valid for the Newtonian case, i.e. for
2nd degree polynomial (see for example \mycite{mcdonnell} ), but
this obviously produces incorrect results, except at the first order
in $\gamma$.

There are no (as far I know) examples of analytical calculations with
higher exponents $s$, valid at any order in $\gamma$, and where
the physical conditions at the orbital inversion points are satisfied.

In this article, a method is proposed for the calculation of this
type of integral, with any exponent $s$, on the basis of integration
in the complex plane. As regards the question of the motion inversion
points, it is solved using the so-called \emph{Sturm's method }\mycite{sturm}.
We also discussed the conditions relating to whether the series converges
or not. 

Some details on the deduction of the values of $s$ and $\gamma$
are in \emph{Appendix A}, while the main calculation method is presented
in \emph{Appendix B (}see also \mycite{mark}).

For a complete review of the physics issues related to anomalous perihelion
precession, see for example \mycite{ellis-1,hervik-1,landau2-1,stephani-1}. 

The outline of this work is as follows. In Section III the series
that gives the perihelion shift $\Delta\phi$ is deduced in general
form. In Section IV the issue of series convergence is discussed;
in Section V the energy and angular momentum formulas are deduced
as a function of the geometric parameters of the orbit. Section VI-VII
deals with the case $s=2,3$, with a free $\gamma$, while section
VIII uses the right $\gamma$ for the GR. In this section, which is
the most important of the work, the result is compared with the works
present in the literature. In Section IX we apply the mehod to the
special case of the planet Mercury. Section X presents the case $s=4$,
as an example, while in section XI the same result is compared with
the value obtained by numerically calculating the integral. Finally,
Section XII presents first-order and second-order results in $\gamma$
for all exponents s between 2 and 7.

\section{Perihelion precession}

In Newton's theory, motion in a central field can be described by
an effective potential \emph{$V_{0}(r)$, }which is obtained by adding
two terms: the gravitational potential $-\frac{\alpha}{r}$, where
$\alpha=M\cdot G$ and where M is the mass of the body that generates
the gravitation force, and the centrifugal potential $\frac{1}{2}\omega^{2}r^{2}=\frac{L^{2}}{2r^{2}}$,
where $L=r^{2}\omega$ is angular momentum per unit mass (constant
of motion).

The relativistic version, $V(r),$ valid for spherically symmetric
mass, contains an additional term expressed as $-\frac{1}{r^{3}}$.
Therefore, as a general case, we assume
\begin{equation}
V(r)=-\frac{\alpha}{r}+\frac{L^{2}}{2r^{2}}+\frac{\gamma}{r^{s}}
\end{equation}

For Schwarzschild's metric, $\gamma$ is $-\frac{\alpha L^{2}}{c^{2}}$.
For a derivation of this formula in General Relativity (GR), see Appendix
\ref{sec:Il-potenziale-efficace}. We use this value of $\gamma$
and the exponent $s=3$ to compare our results to those presented
in the literature.

During motion, the value of $r$ varies between a minimum value $r=r_{1}$
and a maximum value $r=r_{2}$. In the Newtonian case, the orbit closes
with each revolution. However, the presence of an attractive term
such as $-\frac{1}{r^{3}}$ causes the planet to get a slightly closer
to the Sun, and therefore $r_{1}$ decreases. 

The result is that the perihelion advances by an angular amount $\Delta\phi\ne0(mod\ 2\pi)$
per period (see Fig. \ref{fig:prec-spostang}).
\begin{center}
\begin{figure}
\centering{}\includegraphics[scale=0.3]{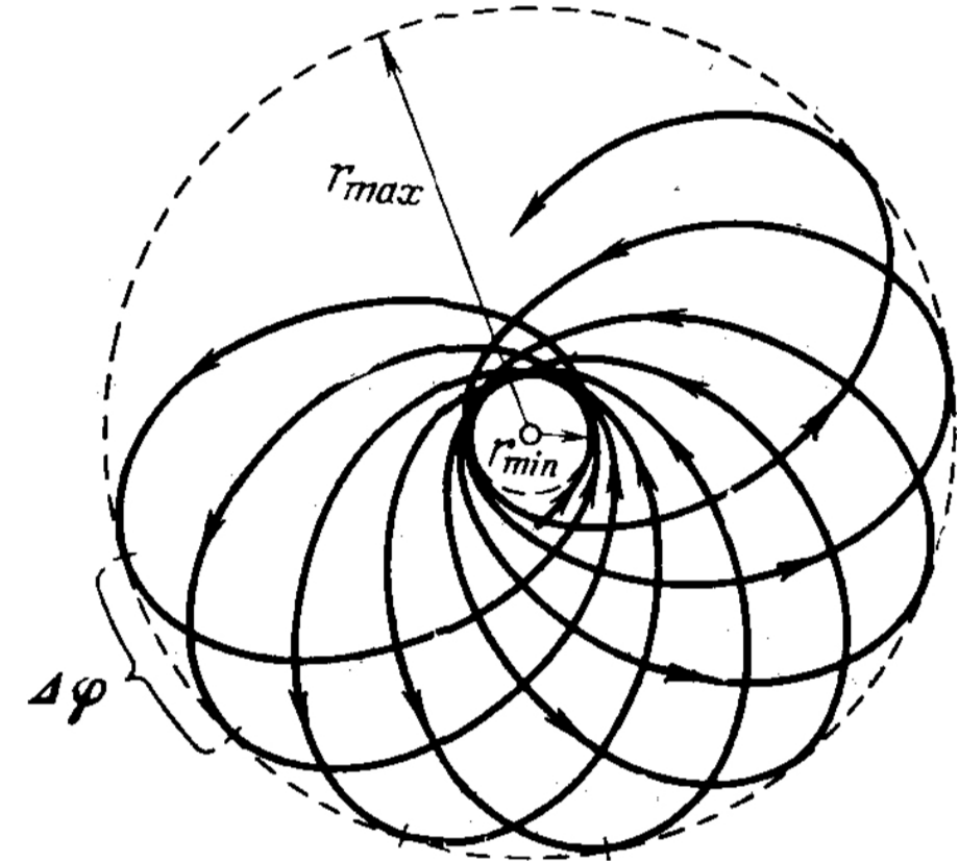}\caption{Perihelion shift of the orbital major axis}
\label{fig:prec-spostang}
\end{figure}
\par\end{center}

\section{Perturbative calculus}

The starting point is to decompose the potential into two parts:
\begin{equation}
V=V_{0}+\delta V,
\end{equation}
treating $\delta V=\frac{\gamma}{r^{s}}$ as a small perturbation.

Using the conservation of mechanical energy, the perihelion shift
$\Delta\phi$ can be obtained by observing that $d\phi=\frac{L}{r^{2}}dt$
and that $dt=\frac{1}{\dot{r}}dr$ (see Appendix A for the definition
of $E_{0}$):

\begin{equation}
\Delta\phi={\displaystyle \oint}_{orbit}\frac{L}{r^{2}}dt={\displaystyle 2\int_{r_{1}}^{r_{2}}}\frac{\frac{L}{r^{2}}}{\sqrt{2(E_{0}-V(r))}}dr
\end{equation}

or:
\begin{equation}
\Delta\phi={\displaystyle 2L\int_{r_{1}}^{r_{2}}}\frac{1}{r^{2}\sqrt{2E_{0}+\frac{2\alpha}{r}-\frac{L^{2}}{r^{2}}-\frac{2\gamma}{r^{s}}}}dr\label{eq:deltaphi_def}
\end{equation}

where $r_{1,2}$ are the inversion points of the motion (positives
zeros of the radicand function).

Moving $-\frac{L^{2}}{r^{2}}$ out of the square root, the integral
\eqref{eq:deltaphi_def} can be written in the form:
\begin{equation}
\Delta\phi=\frac{1}{i}{\displaystyle \oint_{orbit}}\frac{\left(1-A\cdot r-B\cdot r^{2}-\frac{C}{r^{s-2}}\right)^{-\frac{1}{2}}}{r}dr\label{eq:deltaphi_ABC}
\end{equation}

where
\begin{equation}
A=\frac{2\alpha}{L^{2}},\ B=\frac{2E_{0}}{L^{2}},\ C=-\frac{2\gamma}{L^{2}}\label{eq:ABCdef}
\end{equation}

Expanding \eqref{eq:deltaphi_ABC} in powers of C, we have:

\begin{equation}
\Delta\phi=\sum_{n\ge0}\binom{-\frac{1}{2}}{n}(\frac{2\gamma}{L^{2}})^{n}\frac{1}{i}\oint\frac{(1-A\cdot r-B\cdot r^{2})^{-n-\frac{1}{2}}}{r^{1+n(s-2)}}\,dr
\end{equation}

We calculate the involved integral along a path in the complex pane
$r$ that goes around the cut between the points $r_{1}$ ed $r_{2}$
in which the radicand of \eqref{eq:deltaphi_def} vanishes. To do
this, the residues in $r=0$ and $r=\infty$ are required (see Appendix
\ref{sec:Nota-sul-metodo-dei-residui}).

\begin{figure}[h]
\centering{}\centering\includegraphics[scale=0.2]{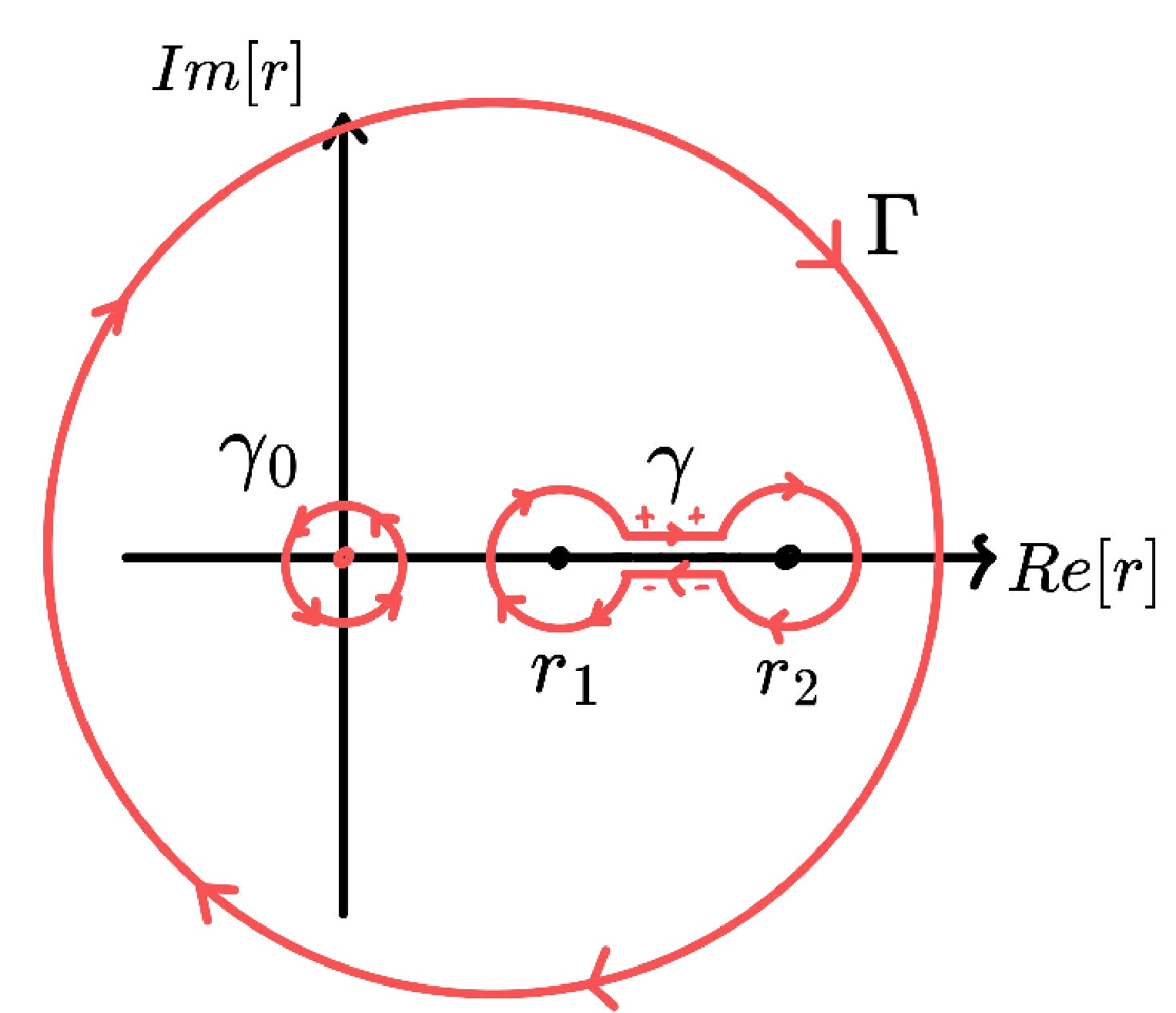}\caption{Paths in the complex plane.}
\label{fig:ossodicane}
\end{figure}

The residue in $r=\infty$ vanishes because the integrating function
become $\frac{r^{-2n-1}}{r^{1+n(s-2)}}=\frac{1}{r^{n\cdot s+2}}$,
that is, $\frac{1}{r^{k}}$, with $k\ge2$.

For the contribution at pole $r=0$, we expand the denominator of
the integrating function in powers of $r$, selecting the term containing
$r^{n(s-2)}$, to obtain the $\sim\frac{1}{r}$ behaviour:

\begin{equation}
(1-A\cdot r-B\cdot r^{2})^{-n-\frac{1}{2}}=\cdots+\text{\fbox{\ensuremath{q_{n}^{s}}}}\cdot r^{n(s-2)}+\cdots
\end{equation}
The coefficient is as follows:

\begin{multline}
{\scriptsize q_{n}^{s}(A,B)=\sum_{p=0}^{\lfloor\frac{n(s-2)}{2}\rfloor}\binom{-n-1/2}{n(s-2)-p}\cdots}\\
\cdots\binom{(ns-2)-p}{p}(-1)^{n(s-2)-p}A^{n(s-2)-2p}B^{p}\label{eq:qns}
\end{multline}

{\footnotesize}{\footnotesize\par}

Multiplying by $2\pi i$, we finally have:

\begin{equation}
\text{\fbox{\fbox{\ensuremath{\Delta\phi}=2\ensuremath{\pi\sum_{n=0}^{\infty}\binom{-\frac{1}{2}}{n}}(\ensuremath{\frac{2\gamma}{L^{2}})^{n}q_{n}^{s}}(\ensuremath{\frac{2\alpha}{L^{2}}},\ensuremath{\frac{2E}{L^{2}}})}}}\label{eq:deltaphi_serie}
\end{equation}

Note that coeffients $q_{n}^{s}$, in \eqref{eq:qns}, for $s=0,1$,
are all zero, as $n$ varies, except for the first coefficient, which
is equal to 1. This finding is in agreement with the theory. In fact:
for $s=0,1$ the perturbation $\frac{\gamma}{r^{s}}$ can be reassorbed
in the case where $\gamma=0$, modifying $E_{0}$ or $\alpha$, so
$\Delta\phi=2\pi=0\ (mod\ 2\pi)$.

In the following we consider only the interesting cases, i.e. $s\ge2$,
computing the formula \eqref{eq:deltaphi_serie} for various powers
$s$, with the help of a symbolic CAS.

\section{Convergence}

First let us consider the asymptotic behavior of the series \eqref{eq:deltaphi_serie}.

Ignoring numerical factors, for $n\to\infty$ we have:
\begin{equation}
\Delta\phi_{n}\sim\left(\frac{\gamma}{L^{2}}\right)^{n}\left(\frac{\alpha}{L^{2}}\right)^{n(s-2)}\sim\left(\frac{\gamma\alpha^{(s-2)}}{(L^{2})^{(s-1)}}\right)^{n}
\end{equation}
In Newtonian limit $L^{2}\approx\alpha\cdot p$, we obtains $\Delta\phi_{n}\sim(\frac{\gamma}{\alpha p^{s-1}})^{n}$,
where $p$ is the so-called \emph{semi-latus rectum, }defined through
the inversion points $r_{1}$ (perihelion) ed $r_{2}$ (aphelion),
and the orbital eccentricity $\epsilon$ with the formulas:
\begin{equation}
r_{1}=\frac{p}{1+\epsilon},\quad r_{2}=\frac{p}{1-\epsilon}\label{eq:semilatus}
\end{equation}

The ratio between two consecutive terms, $\rho=\frac{\gamma}{\alpha p^{s-1}}$,
is dimensionless, because $\gamma/r^{s}$ and $\alpha/r$ must have
the same dimension, so $[\gamma]=[\alpha p^{(s-1)}]$.

Note that the convergence becomes worse for small $\alpha$, unless
$\gamma$ is itself an infinitesimal of the same order, as in GR,
or higher. Divergent values of $\Delta\phi$ mean that, essentially,
the orbit does not close. This was to be expected, because without
the Newtonian part of the potential, one cannot even guarantee that
the orbit is periodic.

Naturally, the series \eqref{eq:deltaphi_serie} converges only under
appropriate conditions on $\gamma,L^{2},\alpha$. Owing to the complexity
of the formulas involved, it is not possible to obtain a set of relations
in closed form, that are valid for each $s$. 

The starting point is the polynomial $P(r,s)$, which is obtained
by multiplying the polynomial in the radicand of \eqref{eq:deltaphi_def}
by $r^{s}$:
\begin{center}
$P(r,s)=$$-\gamma+E_0 r^s-\frac{1}{2} L^2 r^{s-2}+\alpha  r^{s-1}$ \eqnum \label{eq:Prs}
\par\end{center}

In the simplest case, i.e. $s=2$, the problem can be solved easily. 

We need to determine the conditions so that the following polynomial:
\begin{center}
$P(r)=$$-\gamma+E_0 r^2-\frac{L^2}{2}+\alpha  r$ \eqnum
\par\end{center}

has 2 positive zeros. 

Applying \emph{Descartes' rule} of signs,we have:
\begin{equation}
E_{0}<0,\quad\gamma+\frac{L^{2}}{2}>0\label{eq:convergenza-s=00003D2}
\end{equation}

A possible technique, valid for every s, is the so-called \emph{Sturm's
sequence:}

\begin{equation}
\begin{cases}
P_{1}(x)=P(x),\,P_{2}=P'(x)\\
P_{i}=-P_{i-2}\pmod{P_{i-1}} & i>2
\end{cases}\label{eq:sturm-formulas}
\end{equation}

It produces a certain number of inequalities, to be solved. 

For $s=3$, the polynomial is:
\begin{center}
$P(r)=$$-\gamma+E_0 r^3-\frac{L^2 r}{2}+\alpha  r^2$ \eqnum
\par\end{center}

and the request is three positive zeros. Defining $S(r)$ as the number
of sign changes in the sequence, the number of distinct positive roots
is given by $S(0)-S(\infty)$. If we want 3 zeros, we need to impose
that $S(\infty)=0$, i.e. no sign changes, and thant $S(0)=3$, i.e.
alternate signs.

In our case, the sequence \eqref{eq:sturm-formulas} is as follows:
\begin{center}
$\left( \begin{array}{c}  -\gamma +E_0 r^3-\frac{L^2 r}{2}+\alpha   r^2\\  r \left(2 \alpha  +3 E_0 r\right)-\frac{L^2}{2} \\ \frac{18 \gamma  E_0+6 E_0 L^2 r-\alpha   L^2+4\alpha  ^2 r}{18 E_0} \\  \frac{9 E_0 \left(-108\gamma ^2 E_0{}^2+2 E_0 \left(L^6+18 \alpha   \gamma L^2\right)+\alpha  ^2 \left(16 \alpha   \gamma+L^4\right)\right)}{4 \left(2 \alpha  ^2+3 E_0L^2\right){}^2}\\ \end{array} \right)$ \eqnum
\par\end{center}

For $S(0)$, it yields:
\begin{center}
$\left( \begin{array}{c} -\gamma  \\  -\frac{L^2}{2} \\  \frac{18 \gamma E_0-\alpha   L^2}{18 E_0} \\  \frac{9 E_0\left(-108 \gamma ^2 E_0{}^2+2 E_0 \left(L^6+18\alpha   \gamma  L^2\right)+\alpha  ^2 \left(16 \alpha   \gamma+L^4\right)\right)}{4 \left(2 \alpha  ^2+3 E_0L^2\right){}^2}\\ \end{array} \right)$ \eqnum
\par\end{center}

The signs must be $\{+,-,+,-\}$. This produces four inequalities.
Assuming the Newtonian limit, and quasi-circular orbits, the first
three are easy to solve, resulting in the following:
\begin{equation}
L^{2}>0,\ \gamma<0,\ E_{0}<0,\ 9|\gamma|<\alpha p^{2}
\end{equation}

The 4th condition, $P_{4}(0)<0$, which is a second-degree inequality
in $\gamma$, is satisfied in the interval $\gamma_{1}<\gamma<\gamma_{2}$,
where $\gamma_{1,2}$ are as follows:
\begin{center}
{\footnotesize $(\frac{4 \alpha ^3-\sqrt{2} \sqrt{\left(2 \alpha  ^2+3 E_0 L^2\right){}^3}+9\alpha   E_0 L^2}{54 E_0{}^2},\frac{4 \alpha ^3+\sqrt{2} \sqrt{\left(2 \alpha  ^2+3 E_0 L^2\right){}^3}+9\alpha   E_0 L^2}{54E_0{}^2})$}{\footnotesize\par}
\par\end{center}

In the Newtonian limit, they become:
\begin{center}
$(\gamma \to{}-\frac{2 \alpha   p^2}{27},\gamma \to{}0)$ \eqnum
\par\end{center}

Therefore, we have $\frac{27}{2}|\gamma|<\alpha p^{2}$.

In $S(\infty$), keeping the leading term in the limit $r\to\infty$,
the sequence is as follows:
\begin{equation}
\{E_{0}\cdot r^{3},3E_{0}\cdot r^{2},\frac{2}{9}\frac{\alpha^{2}}{E_{0}}r,P_{4}(0)\}
\end{equation}
Imposing that all terms are negative, we find no new constraints,
so the final solution is as follows:
\begin{equation}
E_{0}<0,\ \gamma<0,\ \frac{27}{2}|\gamma|<\alpha p^{2}
\end{equation}
This result can be validated via direct numerical calculus.

Limiting to quasicircular orbit ($\epsilon\to0$), the ratios between
the successive terms of \eqref{eq:deltaphi_serie}, in the range $n\in(5,10,15,20,\dots)$,
are:
\begin{center}
          \begin{tabular}{cc}          \noalign{\arrayrulewidth=8mm}\hline$n$ & $\rho_{n}$\\\hline \hline           $5$ & $-\frac{11.3333\gamma }{\alpha   p^2}$\\           $10$ & $-\frac{12.2975 \gamma}{\alpha   p^2}$\\           $15$ & $-\frac{12.668 \gamma }{\alpha p^2}$\\           $20$ & $-\frac{12.8639 \gamma }{\alpha  p^2}$\\           $25$ & $-\frac{12.9852 \gamma }{\alpha  p^2}$\\\noalign{\arrayrulewidth=8mm}\hline          \noalign{\arrayrulewidth=8mm}\hline           \end{tabular}                   \eqnum
\par\end{center}

They are of the form $K_{n}\frac{|\gamma|}{\alpha p^{2}},$ with $K_{25}\approx12.9$,
slowly growing, and compatible with the theoretical limit value:
\begin{equation}
K=\lim_{n\to\infty}K_{n}=\frac{27}{2}=13.5
\end{equation}

The procedure used is not based on any specific property of the exponent
$s=3$, so we can reasonably hypothesize that the limit ratio $\rho_{\infty}$
is of the form:
\begin{equation}
\rho_{\infty}=\lim_{n\to\infty}\frac{|\Delta\phi_{n+1}|}{|\Delta\phi_{n}|}=K\cdot\frac{|\gamma|}{\alpha p^{(s-1)}}\label{eq:rhoinf}
\end{equation}

where $K$ is a dimensionless constant, depending on $s$ and $\epsilon$. 

We therefore assume the standard convergence requirement $\rho_{\infty}<1$,
i.e. 
\begin{equation}
|\gamma|<\frac{1}{K}\alpha p^{s-1}
\end{equation}

\section{Physical parameters}

Let us now determine the relationship between the physical quantities
$E_{0},L$ and the inversion points $r_{1}$(perihelion) ed $r_{2}$
(aphelion), i.e. the points of minimum and maximum distance to the
Sun.

We describe the shape of the orbit with two orbital empirical parameters
$\epsilon,p$, defined in \eqref{eq:semilatus}. When $\gamma$ is
not small, since we have no longer elliptical orbits, these parameters
lose part of their original meaning. What can reasonably be assumed
is that we are dealing with closed orbits that rigidly rotate by a
certain amount $\Delta\phi\mod{2\pi}$, at every revolution.

The equation to solve, for $s\ge2$, using the same notation as in
\eqref{eq:deltaphi_ABC}, is
\begin{equation}
1-A\cdot r-B\cdot r^{2}-\frac{C}{r^{s-2}}=0
\end{equation}

Multiplying by $r^{s-2}$ and reording, we have:
\begin{equation}
P(r)=r^{s}+\frac{\alpha}{E_{0}}r^{s-1}-\frac{L^{2}}{2E_{0}}r^{s-2}-\frac{\gamma}{E_{0}}=0
\end{equation}

Descartes' rule allows us to determine the maximum number of positive
real solutions, and negative ones, by counting the sign changes in
the coefficients of $P(r)$ and, respectively, in $P(-r)$.

In the case $s=2$, we have two positive solutions if the conditions
\eqref{eq:convergenza-s=00003D2} hold.

For $s\ge3$, from a brief analysis, we realize that the positive
zeros can be 3 or 1. With respect to negative zeros, for odd $s$,
there are none; for even $s$, there is only one. The other zeros,
up to the total of $s$, are complex conjugates.

From a physical point of view, we are only interested in the two positive
zeros $r_{1,2}$ which, within the limit $C\to0$, become the inversion
points of Newtonian motion, whereas the third positive zero $r_{3}$,
if there is, simply tends to zero.

The quantities $E_{0}$ ed $L$ can be determined by eliminating $r_{1}$
ed $r_{2}$ from the $2\times2$ system of equations:
\begin{equation}
\begin{cases}
1-A\cdot r_{1}-B\cdot r_{1}^{2} & =\frac{C}{r_{1}^{s-2}}\\
1-A\cdot r_{2}-B\cdot r_{2}^{2} & =\frac{C}{r_{2}^{s-2}}
\end{cases}\label{eq:sysABC}
\end{equation}

where $A,B,C$ are defined in \eqref{eq:ABCdef}, selecting the two
zeros tending to Newtonian values, when $C$ vanishes.

Once we have found $E_{0},L$ in terms of $r_{1,2}$, we can eliminate
the latter in favor of $\epsilon,p$, determining the functions $E_{0}(p,\epsilon)$
and $L(p,\epsilon)$. If $\gamma$ is simply a constant parameter,
the solutions, for some $s$, are as follows:
\begin{center}
           \begin{tabular}{ccc}         \noalign{\arrayrulewidth=8mm}\hline$s$ & $E_0$ &$L^2$\\\hline \hline           $2$ & $\frac{\alpha  \left(\epsilon ^2-1\right)}{2 p}$ & $\alpha   p-2 \gamma$\\      $3$ & $\frac{\left(\epsilon ^2-1\right) \left(-\gamma \epsilon ^2+\gamma +\alpha   p^2\right)}{2 p^3}$ & $\alpha  p-\frac{\gamma  \left(\epsilon ^2+3\right)}{p}$\\           $4$& $\frac{\left(\epsilon ^2-1\right) \left(\alpha   p^3-2 \gamma \left(\epsilon ^2-1\right)\right)}{2 p^4}$ & $\alpha   p-\frac{4\gamma  \left(\epsilon ^2+1\right)}{p^2}$\\           $5$ &$\frac{\left(\epsilon ^2-1\right) \left(\alpha   p^4-\gamma \left(\epsilon ^4+2 \epsilon ^2-3\right)\right)}{2 p^5}$ & $\alpha p-\frac{\gamma  \left(\epsilon ^4+10 \epsilon^2+5\right)}{p^3}$\\           $6$ & $\frac{\left(\epsilon^2-1\right) \left(\alpha   p^5-4 \gamma  \left(\epsilon^4-1\right)\right)}{2 p^6}$ & $\alpha   p-\frac{2 \gamma  \left(3\epsilon ^4+10 \epsilon^2+3\right)}{p^4}$\\\noalign{\arrayrulewidth=8mm}\hline          \noalign{\arrayrulewidth=8mm}\hline           \end{tabular}                  \eqnum \label{eq:EL}
\par\end{center}

Turning off the perturbation, $\gamma\to0$, they tend to the Newtonian
limit: 
\begin{center}
$E_0=$$\frac{\alpha   (\epsilon -1) (\epsilon +1)}{2p}$, \quad
$L^2=$$\alpha  p$ \eqnum
\par\end{center}

\section{The s=2 case}

Let us now consider the case where $s=2$, with $\gamma$ as the free
parameter.

This is a somewhat special case, because the calculation of the integral
\eqref{eq:deltaphi_def} is easily obtained by a simple substitution
of the parameters. A term like $\frac{\gamma}{r^{2}}$ can, in fact,
be reabsorbed in the centrifugal term $\frac{L^{2}}{2r^{2}}$, rewriting
it as $\frac{\tilde{L}^{2}}{2r^{2}}$, with:
\begin{equation}
\tilde{L}^{2}=L^{2}+2\gamma
\end{equation}

Taking $\frac{\tilde{L}^{2}}{2r^{2}}$ outside the radical in the
integrand, we get:
\begin{equation}
\Delta\phi=\frac{L}{\tilde{L}}\Delta\phi_{\gamma=0}=\frac{L}{\tilde{L}}2\pi=\frac{2\pi}{\left(1+\frac{2\gamma}{L^{2}}\right)^{\frac{1}{2}}}
\end{equation}

In its simplicity, this case gives us a first chance to verify the
correctness of the formal series \eqref{eq:deltaphi_serie}. In fact,
since all the functions $q_{n}^{2}(a,b)$ in \eqref{eq:qns} are equal
to 1, the series \eqref{eq:deltaphi_serie} adds up easily:
\begin{equation}
\Delta\phi=2\pi\sum_{n=0}^{\infty}\binom{-\frac{1}{2}}{n}(\frac{2\gamma}{L^{2}})^{n}=\frac{2\pi}{\left(1+\frac{2\gamma}{L^{2}}\right)^{\frac{1}{2}}}\label{eq:deltaphi2}
\end{equation}

Note that, according to the convergence condition \eqref{eq:convergenza-s=00003D2},
the term under the root must be positive.

Expanding in powers of $\gamma$, we have:
\begin{center}
$\Delta\phi=$$2 \pi -\frac{2 \pi\gamma }{L^2}+\frac{3 \pi  \gamma ^2}{L^4}-\frac{5 \pi  \gamma^3}{L^6}+\frac{35 \pi  \gamma ^4}{4 L^8}-\frac{63 \pi  \gamma^5}{4 L^{10}}+O\left(\gamma^6\right)$ \eqnum
\par\end{center}

The correction term $\Delta\phi_{1}$ agrees with what Landau reported
in \citep[p.40]{landau1}, where $\gamma$ is $\beta$.

\section{The s=3 case}

Let us now consider the case where $s=3$, with $\gamma$ as the free
parameter. This is the first non-trivial case, because the additional
term cannot be reabsorbed simply by replacing the parameters, as done
in the previous paragraph.

For $s=3$, the first 6 functions $q_{n}^{3}(a,b)$ are as follows:
\begin{center}
$\left( \begin{array}{cc} q_0 & 1 \\  q_1 & \frac{3 a}{2} \\  q_2 & \frac{5}{8} \left(7a^2+4 b\right) \\  q_3 & \frac{21}{16} a \left(11 a^2+12 b\right)\\  q_4 & \frac{99}{128} \left(65 a^4+104 a^2 b+16 b^2\right)\\  q_5 & \frac{143}{256} \left(323 a^5+680 a^3 b+240 a b^2\right)\\ \end{array} \right)$ \eqnum
\par\end{center}

The first terms of the $\Delta\phi$ series are, in order:
\begin{center}
{\footnotesize           \begin{tabular}{cc}          \noalign{\arrayrulewidth=8mm}\hline$\Delta \phi _0$ & $2\pi$\\           $\Delta \phi _1$ & $-\frac{6 \pi  \alpha  \gamma }{L^4}$\\           $\Delta \phi _2$ & $\frac{15 \pi \gamma ^2 E_0}{L^6}+\frac{105 \pi  \alpha  ^2 \gamma ^2}{2L^8}$\\           $\Delta \phi _3$ & $-\frac{315 \pi  \alpha  \gamma ^3 E_0}{L^{10}}-\frac{1155 \pi  \alpha  ^3 \gamma^3}{2 L^{12}}$\\           $\Delta \phi _4$ & $\frac{45045 \pi \alpha  ^2 \gamma ^4 E_0}{8 L^{14}}+\frac{3465 \pi  \gamma^4 E_0{}^2}{8 L^{12}}+\frac{225225 \pi  \alpha  ^4 \gamma^4}{32 L^{16}}$\\\noalign{\arrayrulewidth=8mm}\hline          \noalign{\arrayrulewidth=8mm}\hline           \end{tabular}                   \eqnum \label{eq:deltafi3}}{\footnotesize\par}
\par\end{center}

For $E_{0},L$ the substitutions rules found from eq. (\ref{eq:EL})
are as follows:
\begin{center}
$(E_0\to{}\frac{\left(\epsilon ^2-1\right) \left(-\gamma  \epsilon^2+\gamma +\alpha   p^2\right)}{2 p^3},L^2\to{} \alpha  p-\frac{\gamma  \left(\epsilon^2+3\right)}{p})$ \eqnum
\par\end{center}

We have:
\begin{center}
{\small           \begin{tabular}{cc}          \noalign{\arrayrulewidth=8mm}\hline$\Delta \phi _0$ & $2\pi$\\           $\Delta \phi _1$ & $-\frac{6 \pi  \alpha  \gamma  p^2}{\left(\alpha   p^2-\gamma  \left(\epsilon^2+3\right)\right)^2}$\\           $\Delta \phi _2$ & $\frac{15\pi  \gamma ^2 \left(\gamma ^2 \left(\epsilon ^2-1\right)^2\left(\epsilon ^2+3\right)+\alpha  ^2 p^4 \left(\epsilon^2+6\right)-2 \alpha   \gamma  p^2 \left(\epsilon^4-1\right)\right)}{2 \left(\alpha   p^2-\gamma  \left(\epsilon^2+3\right)\right)^4}$\\           $\Delta \phi _3$ &$-\frac{105 \pi  \alpha   \gamma ^3 p^2 \left(3 \gamma ^2\left(\epsilon ^2-1\right)^2 \left(\epsilon ^2+3\right)+\alpha  ^2p^4 \left(3 \epsilon ^2+8\right)-6 \alpha   \gamma  p^2\left(\epsilon ^4-1\right)\right)}{2 \left(\alpha   p^2-\gamma \left(\epsilon^2+3\right)\right)^6}$\\\noalign{\arrayrulewidth=8mm}\hline     \noalign{\arrayrulewidth=8mm}\hline           \end{tabular}             } \eqnum
\par\end{center}

Even in this case, the correction term $\Delta\phi_{1}$ agrees with
what Landau reported in \cite[p.40]{landau1}.

\section{The GR case}

To check the validity of the method, we compare it with the GR results
reported in the literature, where $\delta V=\frac{-\alpha L^{2}}{r^{3}}$.

In GR $\gamma$ is constant of motion, but it is not <<free>>: it
depends on $L$, and this must be considered when solving the system
\eqref{eq:sysABC} for $E_{0}$ and $L$. Using $\gamma=-\alpha L^{2}$
and $s=3$, the result is as follows:
\begin{center}
$(E_0\to{} \frac{\alpha   \left(\epsilon ^2-1\right) (p-4 \alpha  )}{2 p\left(p-\alpha   \left(\epsilon ^2+3\right)\right)},L^2\to{}\frac{\alpha   p^2}{p-\alpha   \left(\epsilon^2+3\right)})$ \eqnum \label{eq:EL_GR}
\par\end{center}

These values tend to Newtonian values, when $\alpha\to0$.

Substituting (\ref{eq:EL_GR}) in eq. (\ref{eq:deltafi3}), we obtain,
up to the 3rd order:
\begin{center}
          \begin{tabular}{cc}          \noalign{\arrayrulewidth=8mm}\hline$\Delta \phi _0$ & $2\pi$\\           $\Delta \phi _1$ & $\frac{6 \pi  \alpha  \left(p-\alpha   \left(\epsilon ^2+3\right)\right)}{p^2}$\\     $\Delta \phi _2$ & $\frac{15 \pi  \alpha  ^2 \left(7 \alpha ^2 \left(\epsilon ^2+3\right)^2+p^2 \left(\epsilon ^2+6\right)-2\alpha   p \left(9 \epsilon ^2+19\right)\right)}{2 p^4}$\\       $\Delta \phi _3$ & $\frac{105 \pi  \alpha  ^3 \left(p-\alpha  \left(\epsilon ^2+3\right)\right) \left(11 \left(p-\alpha  \left(\epsilon ^2+3\right)\right)^2+3 p \left(\epsilon^2-1\right) (p-4 \alpha  )\right)}{2p^6}$\\\noalign{\arrayrulewidth=8mm}\hline          \noalign{\arrayrulewidth=8mm}\hline           \end{tabular}                   \eqnum
\par\end{center}

Introducing the new variable $\zeta=\frac{2\alpha}{p}$, the last
series can be written as follows:
\begin{center}
$\Delta \phi=$
$2 \pi +3 \pi  \zeta+\frac{3}{8} \pi  \zeta ^2 \left(\epsilon^2+18\right)+\frac{45}{16} \pi  \zeta ^3 \left(\epsilon^2+6\right)+\frac{105}{512} \pi  \zeta ^4 \left(\epsilon ^4+72\epsilon ^2+216\right)+\frac{567 \pi  \zeta ^5 \left(5 \epsilon^4+120 \epsilon ^2+216\right)}{1024}+\frac{231 \pi  \zeta ^6\left(5 \epsilon ^6+810 \epsilon ^4+9720 \epsilon^2+11664\right)}{8192}+O\left(\zeta^7\right)$ \eqnum
\par\end{center}

This result agrees with that reported by \mycite{poveda-1}, with
the exception of 5th term.

The power expansion in $\alpha/r$, in the limit $\epsilon\to0$,
which is useful for quasicircular orbits of radius $r$, is also interesting:
\begin{center}
$\Delta \phi=2\pi$(
$1+\frac{3\alpha  }{r}+\frac{27 \alpha  ^2}{2 r^2}+\frac{135 \alpha  ^3}{2r^3}+\frac{2835 \alpha  ^4}{8 r^4}+\frac{15309 \alpha  ^5}{8r^5}+O\left(\alpha  ^6\right)$
) \eqnum
\par\end{center}

All the coefficients found agree with \mycite{sjwalters-1}, who calculated
them with the help of the elliptic function of the first kind $K(x)$.
However, they do not agree with those of \mycite{deliseo-1}.

\section{Mercury's precession}

For the planet Mercury, the measured precession is $43.1\pm0.5$ seconds
for a century.

Assuming that $\zeta=\frac{2\alpha}{p}=$$5.35\times 10^{-8}$,
the corrections, which are calculated with \eqref{eq:deltaphi_serie},
are as follows:
\begin{center}
          \begin{tabular}{cc}          \noalign{\arrayrulewidth=8mm}\hline$order$ &$\frac{arcsecs}{century}$\\\hline \hline          $\Delta \phi _1$ & $43.1939$\\           $\Delta \phi _2$ &$8.72907425318373\times 10^{ -6}$\\          $\Delta \phi _3$ & $2.1989864422543495\times 10^{-12}$\\           $\Delta \phi _4$ &$6.124758430483005\times 10^{ -19}$\\          $\Delta \phi _5$ & $1.8069811065834542\times 10^{-25}$\\           $\Delta \phi _6$ &$5.533906576517437\times 10^{ -32}$\\          $\Delta \phi _7$ & $1.739721211748118\times 10^{-38}$\\           $\Delta \phi _8$ &$5.5762541361063566\times 10^{-45}$\\\noalign{\arrayrulewidth=8mm}\hline       \noalign{\arrayrulewidth=8mm}\hline           \end{tabular}               
\eqnum
\par\end{center}

The series converges rapidly: each term is approximately on millionth
of the previous term, so that even the second-order correction extends
well beyond the current measurement capabilities.

\section{The s=4 case}

In this Section we test the formula \eqref{eq:deltaphi_serie} for
$s=4$.

Some $q_{n}^{4}(a,b)$ polynomials are as follows:
\begin{center}
           \begin{tabular}{cc}          \noalign{\arrayrulewidth=8mm}\hline$q_0$ & $1$\\           $q_1$ &$\frac{3}{8} \left(5 a^2+4 b\right)$\\           $q_2$ &$\frac{35}{128} \left(33 a^4+72 a^2 b+16 b^2\right)$\\          $q_3$ & $\frac{231 \left(221 a^6+780 a^4 b+624 a^2 b^2+64b^3\right)}{1024}$\\\noalign{\arrayrulewidth=8mm}\hline          \noalign{\arrayrulewidth=8mm}\hline           \end{tabular}                   \eqnum
\par\end{center}

The first terms of the $\Delta\phi$ series are, as follows:
\begin{center}
{\footnotesize        \begin{tabular}{cc}          \noalign{\arrayrulewidth=8mm}\hline$\Delta \phi _0$ & $2\pi$\\           $\Delta \phi _1$ & $-\frac{6 \pi  \gamma E_0}{L^4}-\frac{15 \pi  \alpha  ^2 \gamma }{L^6}$\\       $\Delta \phi _2$ & $\frac{945 \pi  \alpha  ^2 \gamma ^2E_0}{2 L^{10}}+\frac{105 \pi  \gamma ^2 E_0{}^2}{2L^8}+\frac{3465 \pi  \alpha  ^4 \gamma ^2}{8L^{12}}$\\\noalign{\arrayrulewidth=8mm}\hline          \noalign{\arrayrulewidth=8mm}\hline           \end{tabular}                   \eqnum}{\footnotesize\par}
\par\end{center}

The substitution rules for $E_{0},L$ are:
\begin{center}
$(E_0\to{}\frac{\left(\epsilon ^2-1\right) \left(\alpha   p^3-2 \gamma \left(\epsilon ^2-1\right)\right)}{2 p^4},L^2\to{} \alpha  p-\frac{4 \gamma  \left(\epsilon^2+1\right)}{p^2})$ \eqnum
\par\end{center}

Substituting $E_{0},L$ we obtain rational, rather complicated expressions.

For example, the first-order correction is as follows:
\begin{center}
$\Delta\phi_1=$\break$-\frac{3 \pi  \gamma \left(8 \gamma ^2 \left(\epsilon ^2-1\right)^2 \left(\epsilon^2+1\right)+\alpha  ^2 p^6 \left(\epsilon ^2+4\right)+2 \alpha  \gamma  p^3 \left(-3 \epsilon ^4+2 \epsilon^2+1\right)\right)}{\left(\alpha   p^3-4 \gamma  \left(\epsilon^2+1\right)\right)^3}$ \eqnum
\par\end{center}

The second-order correction is:
\begin{center}
$\Delta\phi_2=$$\frac{105 \pi  \gamma ^2\left(\epsilon ^4+16 \epsilon ^2+16\right)}{8 \alpha  ^2p^6}+O\left(\gamma ^3\right)$ \eqnum
\par\end{center}

and the third-order is:
\begin{center}
$\Delta\phi_3=$$-\frac{1155\gamma ^3 \left(\pi  \left(\epsilon ^6+36 \epsilon ^4+120\epsilon ^2+64\right)\right)}{16 \left(\alpha  ^3p^9\right)}+O\left(\gamma^4\right)$ \eqnum
\par\end{center}

For the case where s=3, in the limit of circular orbits, the ratios
between the successive terms $\rho_{n}$, for $n\in(5,10,15,\dots)$
are as follows:
\begin{center}
$\rho_n=$$(-\frac{26.8333 \gamma }{\alpha   p^3},-\frac{29.1405 \gamma }{\alpha  p^3},-\frac{30.0234 \gamma }{\alpha  p^3})$ \eqnum
\par\end{center}

These results suggest a K of slightly more than 30.

\section{Numerical simulation}

In this $s=4$ simulation, for the parameters A,B, and C we choose
a set of values with no \textquotedbl physical\textquotedbl{} meaning:
\begin{center}
$A=$$2$,$B=$$-1$, $C=$$\frac{1}{500}$ \eqnum
\par\end{center}

\begin{figure}[h]
\centering{}\centering\includegraphics[scale=0.6]{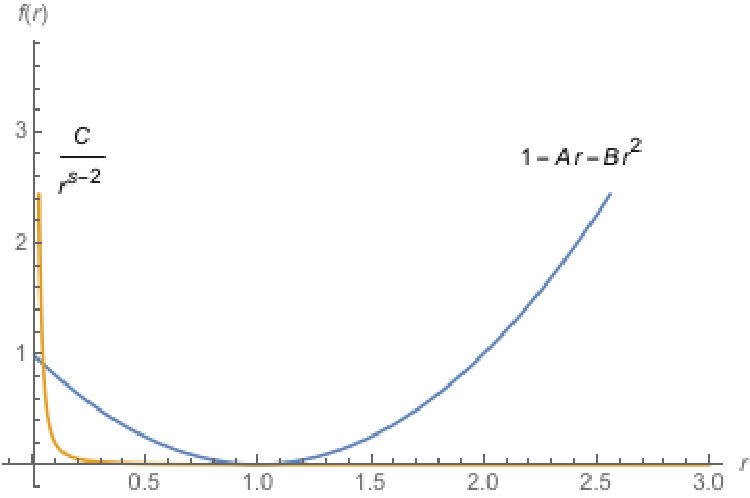}\caption{Inversion points for $s=4$. See eq. \eqref{eq:sysABC}}
\label{fig:plot4}
\end{figure}

The <<non perturbative>> value of $\Delta\phi$ can be obtained
by evaluating the integral in (\ref{eq:deltaphi_ABC}) numerically,
in the interval (see Fig. \eqref{fig:plot4}):
\begin{center}
$r_1=$$0.953077$, $r_2=$$1.04288$ \eqnum
\par\end{center}

With the precision provided by the software routine, one finds the
following:
\begin{center}
$\Delta\phi\approx$ $6.32156$ (rad) 
\par\end{center}

The first terms of the series \eqref{eq:deltaphi_serie} are:
\begin{center}
          \begin{tabular}{cc}          \noalign{\arrayrulewidth=8mm}\hline$n$ & $\Delta \phi_{n}$\\\hline \hline           $0$ & $6.28319$\\        $1$ & $0.0376991$\\           $2$ & $0.000659734$\\          $3$ & $0.0000145142$\\           $4$ &$3.537826027023806\times 10^{-7}$\\\noalign{\arrayrulewidth=8mm}\hline        \noalign{\arrayrulewidth=8mm}\hline           \end{tabular}                 \eqnum
\par\end{center}

obtaining:
\begin{center}
$\Delta \phi=\sum_n \Delta\phi_n=$
$6.32156$ (rad)
\par\end{center}

The difference, which is on the 5th digit after the decimal point,
is probably due more to the numerical integration routine than to
the truncation of the series, given the rapid convergence.

\section{Other results}

It is impossible in a paper to report all the long expressions that
are obtained from the $\Delta\phi$ series, for all the $s$ exponents
and for all the orders in $\gamma$. However, we report some of the
results as an example, for first and second-order.

$\Delta\phi_{1}$ correction, up to exponent $s=7$ (to first order
in $\gamma$):
\begin{center}
          \begin{tabular}{cc}          \noalign{\arrayrulewidth=8mm}\hline$s$ & $\Delta \phi_1$\\\hline \hline           $2$ & $-\frac{2 \pi  \gamma}{\alpha   p}$\\           $3$ & $-\frac{6 \pi  \gamma }{\alpha  p^2}$\\           $4$ & $-\frac{3 \pi  \gamma  \left(\epsilon^2+4\right)}{\alpha   p^3}$\\           $5$ & $-\frac{5 \pi \gamma  \left(3 \epsilon ^2+4\right)}{\alpha   p^4}$\\          $6$ & $-\frac{15 \pi  \gamma  \left(\epsilon ^4+12 \epsilon^2+8\right)}{4 \alpha   p^5}$\\           $7$ & $-\frac{21 \pi \gamma  \left(5 \epsilon ^4+20 \epsilon ^2+8\right)}{4 \alpha  p^6}$\\           $8$ & $-\frac{7 \pi  \gamma  \left(5 \epsilon^6+120 \epsilon ^4+240 \epsilon ^2+64\right)}{8 \alpha  p^7}$\\\noalign{\arrayrulewidth=8mm}\hline          \noalign{\arrayrulewidth=8mm}\hline           \end{tabular}                   \eqnum
\par\end{center}

$\Delta\phi_{2}$ correction (to second order in $\gamma$):
\begin{center}
          \begin{tabular}{cc}          \noalign{\arrayrulewidth=8mm}\hline$s$ & $\Delta \phi_2$\\\hline \hline           $2$ & $\frac{3 \pi  \gamma^2}{\alpha  ^2 p^2}$\\           $3$ & $\frac{15 \pi  \gamma ^2\left(\epsilon ^2+6\right)}{2 \alpha  ^2 p^4}$\\           $4$ &$\frac{105 \pi  \gamma ^2 \left(\epsilon ^4+16 \epsilon^2+16\right)}{8 \alpha  ^2 p^6}$\\           $5$ & $\frac{315 \pi \gamma ^2 \left(\epsilon ^6+30 \epsilon ^4+80 \epsilon^2+32\right)}{16 \alpha  ^2 p^8}$\\           $6$ & $\frac{495 \pi\gamma ^2 \left(7 \epsilon ^8+336 \epsilon ^6+1680 \epsilon^4+1792 \epsilon ^2+384\right)}{128 \alpha  ^2 p^{10}}$\\          $7$ & $\frac{3003 \pi  \gamma ^2 \left(3 \epsilon ^{10}+210\epsilon ^8+1680 \epsilon ^6+3360 \epsilon ^4+1920 \epsilon^2+256\right)}{256 \alpha  ^2p^{12}}$\\\noalign{\arrayrulewidth=8mm}\hline          \noalign{\arrayrulewidth=8mm}\hline           \end{tabular}                   \eqnum
\par\end{center}

\section{Conclusions and outlook}

In this work the problem of the perturbative calculation of the perihelion
shift for non-Newtonian potential $\frac{\gamma}{r^{s}}$ was addressed.
The corrections were obtained, to all orders, by calculating the relevant
integrals in the complex plane, appropriately bypassing the singularities.
The results for case $s=3$ were compared with those presented in
the literature, relating to the Schwarzschild metric, and are in agreement.
The method was finally applied, as case study, to the exponent $s=4$.
For this specific case, the non-perturbative value was also calculated
numerically, obtaining a precision of 2 parts per million.

The use of complex analysis made it possible to obtain a closed formula,
in the form of a power series, valid for all exponents $s$ and any
eccentricity $\epsilon$. Unlike other similar works (for example:
\mycite{hall,mcdonnell}), our result is valid at any order and correctly
takes into account that the physical parameters of the orbit (energy
and angular momentum) no longer have Newtonian values.

Many central-force modifications to gravity can be found in the literature,
all of which can be treated according to the methods developed here. 

An example of these is the contribution to perihelion shift coming
from the high-order multipole expansion of the density mass distribution
$\rho(x)$:
\begin{equation}
V(r)=-\frac{\alpha}{r}\cdot\left(1+\sum_{n\ge2}\frac{q_{n}}{r^{n}}\right)
\end{equation}

where $q_{n}$ are numerical coefficients that encode the non-sphericity
of the central body (See \citet[eq 3.54]{straumann}). The quadrupole
term ($n=2)$ leads to the $\frac{1}{r^{3}}$ contribution, and can
be treated in a standard way with elliptic integrals. The octupole
term ($n=3)$, instead, produces a contribution of the type $\frac{1}{r^{4}}$,
for which the presented method could be usefully employed.

Another possible use of the method is the calculation of the bending
of light by a star (See \cite[eq 3.57]{straumann}): the equations
are the same used for the perihelion's shift, and are reduced to these
by simple reparametrization. Even in this case, the non-sphericity
of the central body (for example, a binary star) could be taken into
account via multipolar expansion.

\textbf{Acknowledgement}: Thanks to Bruno Cocciaro, Pier Franco Nali
and Elio Proietti (ISF), for useful discussions and for pointing out
an inaccuracy in paragraph III.

\textbf{Declarations}: For this work there is no Funding and/or Conflicts
of interests/Competing interests.

\appendix

\section{The effective \foreignlanguage{british}{potential} $V(r)$\protect\label{sec:Il-potenziale-efficace}}

Consider a particle of unit mass $m=1$ moving around a gravitational
centre of mass M. Following the General Relativity, particle's path
is a timelike geodesic $x^{\mu}(\tau)$ in the spacetime, whose metric
is is determined by the gravitational field. Geodesics are described
as stationary points of the functional $\int\mathcal{L}(x^{\mu},\dot{x}^{\mu})d\tau$,
where $\mathrm{\mathcal{L}}=\frac{ds^{2}}{d\tau^{2}}$ is the Lagrangian
and $ds^{2}$ si the quadratic form related to the Schwarzschild metric. 

Using geometric units ($c=1,G=1)$, the Newtonian potential is $-\frac{\alpha}{r}$,
with $\alpha=M$, and the Schwarzschild radius is $r_{S}=2\alpha$,
so we have \mycite{rindler-1}:
\begin{equation}
ds^{2}=(1-\frac{2\alpha}{r})dt^{2}-\frac{1}{(1-\frac{2\alpha}{r})}dr^{2}-r^{2}(d\theta^{2}+\sin^{2}\theta d\phi^{2})
\end{equation}

where $(r,\phi,\theta)$ are the sperical coordinates and $t$ is
the time, as measured by an observer at $r\to\infty$. As is known
from Mechanics, due to the conservation of angular momentum, the orbital
motion occurs entirely in a plane, so we can assume $\theta=\pi/2$
without losing generality. 

Indicating with a dot $(\dot{})$ the first derivative with respect
to the particle's proper time $\tau$, and putting $d\theta=0$ in
the metric, we find \mycite{ellis-1,schutz-1,stephani-1}:
\begin{equation}
\mathrm{\mathcal{L}}=(1-\frac{2\alpha}{r})\cdot\dot{t}{}^{2}-\frac{1}{(1-\frac{2\alpha}{r})}\cdot\dot{r}{}^{2}-r^{2}\dot{\phi}{}^{2}
\end{equation}

The solution $x^{\mu}(\tau)=\left(t(\tau),r(\tau),\phi(\tau)\right)$
is obtained by solving, for $r>r_{s}$, the Lagrange's system of equations:
\begin{equation}
\frac{\partial\mathcal{L}}{\partial x^{\mu}}=\frac{d}{d\tau}\frac{\partial\mathcal{L}}{\partial\dot{x^{\mu}}}
\end{equation}
Since $\mathcal{L}$ does not explicitly depend on either $t$ or
$\phi$, we have two constants of motion, which can be obtained by
differentiating with respect to $\dot{t}$ and $\dot{\phi}$, respectively:
the (relativistic) energy $E$ and the angular momentum $L$:
\begin{equation}
(1-\frac{2\alpha}{r})\cdot\dot{t}=E,\quad r^{2}\dot{\phi}=L
\end{equation}

For a free-falling particle, $d\tau$ coincides with the line element
$ds$ of the metric and the Lagrangian $\mathrm{\mathcal{L}}=\frac{dx_{\mu}dx^{\mu}}{d\tau^{2}}$
is numerically equal to 1. This fact can be exploited to obtain $\dot{r}(\tau)$
without solving the Lagrange equation, simply rewritten as:

\begin{equation}
\dot{r}^{2}=(E^{2}-1)+\frac{2\alpha}{r}-\frac{L^{2}}{r^{2}}+\text{\fbox{\ensuremath{\frac{2\alpha\ensuremath{L^{2}}}{r^{3}}}}}
\end{equation}

The quantity $E^{2}-1$ is itself a constant of motion, and we rename
it $2E_{0}$. This is due to the need to subtract the rest energy
$mc^{2}$ of a unit mass. With this definition, we write:
\begin{equation}
\left(\frac{dr}{d\tau}\right)^{2}=2\left(E_{0}-V(r)\right)
\end{equation}
Apart from the presence of the proper time $\tau$, this equation
corresponds to the motion of a particle of mass $m=1$ and energy
$E_{0}=\frac{\dot{r}^{2}}{2}+V(r)$, in the effective potential:
\begin{equation}
V(r)=-\frac{\alpha}{r}+\frac{L^{2}}{2r^{2}}-\text{\ensuremath{\frac{\alpha L^{2}}{r^{3}}}}
\end{equation}

\section{Note on the residues\protect\label{sec:Nota-sul-metodo-dei-residui}}

The method of residues consists of extending the integration over
closed curves in the complex plane. The method is based on the Cauchy
Theorem which, essentially, states that the integral of a function
on a closed path $\gamma$ is equivalent to the sum of the integrals
made around all the internal isolated singularities in the path; equivalently,
one can use singularities external to the path (including the point
at infinity $r\to\infty$), but change the sign of the result.

The basic fact is that: integrals as $\oint r^{n}dr$, with $n\in\mathcal{Z}$,
on closed anticlockwise curves around ad $r=0$, are all zeros, except
whe $n=-1$:
\begin{equation}
\oint\frac{1}{r}dr=2\pi i
\end{equation}
The other integrals are obtained from these, expanding the integrand
in the generalized Taylor series. 

With respect to non-integer powers, for example square roots, special
attention must be paid to the points where the radicands cancel, bypassing
them with appropriate closed paths. This is precisely our case: we
have two roots, the reversal points of motion, which will be bypassed
with the so-called \textquotedbl bone\textquotedbl{} path.

As an example the method, we demonstrate the formula in \cite[p.167]{landau1}
relating to the radial action in Keplerian motion:
\begin{align}
S_{0} & =2\int_{r_{1}}^{r_{2}}\sqrt{2E_{0}+\frac{2\alpha}{r}-\frac{L^{2}}{r^{2}}}\ dr=\\
= & -2\pi L+\frac{2\pi\alpha}{\sqrt{2|E_{0}|}}\label{eq:landau_s0}
\end{align}

where $r_{1}$ and $r_{2}$ are the inversion points, i.e. the zeros
of the radicand.

$S_{0}$ can be rewritten as an integral over the closed <<bone>>
path in the complex plane surrounding $r_{1}$ e $r_{2}$, choosing
the positive sign for the root on the upper edge of the cut.

In fact, note that:
\begin{equation}
\oint_{\gamma}f(r)dr=2\int_{r_{1}}^{r_{2}}f(r)dr+\oint_{C_{1}}f(r)dr+\oint_{C_{2}}f(r)dr
\end{equation}

and that the integrals on the two small circles $C_{1}$ and $C_{2}$,
centered on the inversion points, vanish when their radius tends to
zero.

According to Cauchy's theorem, which focuses on singularities outside
$\gamma$, $S_{0}$ takes contributions only from $r=0$ and $r=\infty$
(see Fig. \ref{fig:ossodicane}):
\begin{equation}
\oint_{\gamma}=\oint_{\gamma_{0}}+\oint_{\Gamma}
\end{equation}

\textbf{Contribution $r=0$. }If $r$ is small, only the main leading
term $1/r^{2}$ remains in the integral, so:

\begin{equation}
\oint\sqrt{-\frac{L^{2}}{r^{2}}+\cdots}\ dr=i\cdot L\oint\frac{1}{r}\ dr=-2\pi L
\end{equation}

\textbf{Contribution $r=\infty$. }Let us rewrite the integral, as:

$\oint\sqrt{2E_{0}+\frac{2\alpha}{r}-\frac{L^{2}}{r^{2}}}\ dr=\sqrt{2E}\cdot\oint\sqrt{1+\frac{\alpha}{E_{0}\cdot r}-\frac{L^{2}}{2E_{0}r^{2}}}\ dr$

In the binomial expansion $(1+x)^{1/2}=1+\frac{1}{2}x+\dots$, the
only non-zero contribution comes from the term containing $1/r$,
so:

\begin{align*}
\sqrt{2E_{0}}\cdot\oint(1+\frac{\alpha}{2E_{0}\cdot r}+\cdots)\ dr & =\\
\frac{\alpha}{2E_{0}}\sqrt{2E_{0}}\cdot\oint(\frac{1}{r}+\cdots)\ dr & =\frac{2\pi\alpha}{\sqrt{2|E_{0}|}}
\end{align*}

and this proves the Landau's formula \eqref{eq:landau_s0}.

%\nocite{*}


\begin{thebibliography}{0}%
\makeatletter
\providecommand \@ifxundefined [1]{%
 \@ifx{#1\undefined}
}%
\providecommand \@ifnum [1]{%
 \ifnum #1\expandafter \@firstoftwo
 \else \expandafter \@secondoftwo
 \fi
}%
\providecommand \@ifx [1]{%
 \ifx #1\expandafter \@firstoftwo
 \else \expandafter \@secondoftwo
 \fi
}%
\providecommand \natexlab [1]{#1}%
\providecommand \enquote  [1]{``#1''}%
\providecommand \bibnamefont  [1]{#1}%
\providecommand \bibfnamefont [1]{#1}%
\providecommand \citenamefont [1]{#1}%
\providecommand \href@noop [0]{\@secondoftwo}%
\providecommand \href [0]{\begingroup \@sanitize@url \@href}%
\providecommand \@href[1]{\@@startlink{#1}\@@href}%
\providecommand \@@href[1]{\endgroup#1\@@endlink}%
\providecommand \@sanitize@url [0]{\catcode `\\12\catcode `\$12\catcode
  `\&12\catcode `\#12\catcode `\^12\catcode `\_12\catcode `\%12\relax}%
\providecommand \@@startlink[1]{}%
\providecommand \@@endlink[0]{}%
\providecommand \url  [0]{\begingroup\@sanitize@url \@url }%
\providecommand \@url [1]{\endgroup\@href {#1}{\urlprefix }}%
\providecommand \urlprefix  [0]{URL }%
\providecommand \Eprint [0]{\href }%
\providecommand \doibase [0]{https://doi.org/}%
\providecommand \selectlanguage [0]{\@gobble}%
\providecommand \bibinfo  [0]{\@secondoftwo}%
\providecommand \bibfield  [0]{\@secondoftwo}%
\providecommand \translation [1]{[#1]}%
\providecommand \BibitemOpen [0]{}%
\providecommand \bibitemStop [0]{}%
\providecommand \bibitemNoStop [0]{.\EOS\space}%
\providecommand \EOS [0]{\spacefactor3000\relax}%
\providecommand \BibitemShut  [1]{\csname bibitem#1\endcsname}%
\let\auto@bib@innerbib\@empty
%</preamble>
\end{thebibliography}%


\begin{thebibliography}{}
  
\bibitem[Stephani(2000)]{stephani-1}H. Stephani: Relativity: an introdution
to Special and General Relativity, 3ed, Cambridge.
\bibitem[Ellis(2006)]{ellis-1}G. Ellis, R. Williams, Flat and curved
space-times, 2ed, Oxford.
\bibitem[Rindler(2006)]{rindler-1}W. Rindler, Relativity: Special,
General and Cosmological, 2ed, 2006,Oxford.

\bibitem[Landau-Lifshitz(1975)]{landau2-1}L.D. Landau and E. M. Lifshitz,
Teoria dei Campi, vol 2, Editori Riuniti.

\bibitem[Landau-Lifshitz(1976)]{landau1}L. D. Landau and E. M. Lifshitz,
Mechanics, Oxford,1976, 3rd

\bibitem[Ohanan-Ruffini(2000)]{key-14}Ohanian-Ruffini, Gravitation
and space-time, 3ed, 2000,Cambridge.

\bibitem[Schutz(2017)]{schutz-1}B. Schutz, A first course in General
Relativity, 2ed, 2017,Cambridge.


\bibitem[Hervik(2017)]{hervik-1}\O yvind Gr\o n, Sigbj\o rn Hervik,
Einstein\textquoteright s General Theory of Relativity, Springer,
2017.

\bibitem[Carrol(2018)]{carroll}S. Carroll, Spacetime and Geometry:
An Introduction to General Relativity, Cambridge Univ. Press, 2019

\bibitem[Gasperini(2018)]{gasperini}M. Gasperini, Theory of Gravitational
Interactions, Springer, 2013

\bibitem[Wald(1971)]{wald}R. Wald, General Relativity, Chicago Univ.
Press

\bibitem[Moore(1971)]{moore}T.A. Moore, A General Relativity Workbook,
Pomona College, 2013

\bibitem[Walters(2018)]{sjwalters-1}S.J. Walters, A simple exact
series representation for relativistic perihelion advance, R. Astr.
Soc., 2018

\bibitem[Poveda-Mar\`in(2018)]{poveda-1}Poveda, Mar\`in. Perihelion
precession in binary systems: higher order corrections, 2018 http://arxiv.org/abs/1802.03333v1

\bibitem[D'Eliseo(2010)]{deliseo-1} M. D\textquoteright Eliseo. Higher-order
corrections to the relativistic perihelion advance and the mass of
binary pulsar. Astr. and Space. Sc., 2010.

\bibitem[Gine(2005)]{gine-1}Jaume Gin\'e, On the origin of the anomalous
precession of mercury's perihelion, Arxiv, 2005.

\bibitem[Markusevic(1977)]{mark} A.I. Markusevic, Theory of Analytical
Functions: A Brief Course, Mir Publishers, 1977

\bibitem[Adkins(2007)]{mcdonnell} Gregory S. Adkins and J. McDonnel, Orbital precession
due to central-force perturbations, Physical Review D 75, 082001 (2007)

\bibitem[Biswas(2024)]{biswas}Biswas, S.  The Precession of Perihelion
over Modification of Newton Gravity, Physics of the Dark Universe,
43, 101403, (2024)

\bibitem[Hall(2022)]{hall} Michael J. W. Hall, Simple precession
calculation for Mercury: a linearization approach, https://arxiv.org/abs/2206.11617v1,
(2022)

\bibitem[Sturm(2025)]{sturm}Wikipedia, Sturm's method, https://en.wikipedia.org/wiki/Sturm27s\_theorem

\bibitem[Straumann(2004)]{straumann}Straumann, General Relativity
With Applications To Astrophysics, Springer, 2004

\end{thebibliography}
\end{document}